\documentclass[aps,prd,reprint,groupedaddress,nofootinbib,floatfix]{revtex4-2}

\usepackage{amssymb}
\usepackage{amsthm}
\usepackage{mathtools}
\usepackage{enumitem}
\usepackage{graphicx}
\usepackage{hyperref}
\usepackage{cleveref}
\usepackage[dvipsnames]{xcolor}
\usepackage{tikz}
\usetikzlibrary{positioning}
\usepackage{siunitx}
\usepackage{float}
\usepackage[T1]{fontenc}
\usepackage{booktabs}
\usepackage{natbib}

\hypersetup{
	colorlinks=true,
	linkcolor=blue,  
	urlcolor=blue,
	citecolor=blue
}

\definecolor{mylinkcolor}{RGB}{0,0,0}  

\definecolor{lightred}{rgb}{0.884 0.584 0.619}
\definecolor{darkred}{rgb}{0.363 0 0}
\definecolor{lightblue}{rgb}{0.555 0.693 0.804}
\definecolor{darkblue}{rgb}{0 0 0.363}
\definecolor{lightgreen}{rgb}{0.462 0.835 0.462}
\definecolor{darkgreen}{rgb}{0 0.392 0}

\definecolor{lightpink}{rgb}{0.889 0.584 0.72}
\definecolor{darkred2}{rgb}{0.387 0 0.552}

\definecolor{lightblue2}{rgb}{0.493 0.616 0.755}
\definecolor{lightblue3}{rgb}{0.432 0.539 0.706}
\definecolor{pinkmaple}{rgb}{0.878 0.533 0.910}

\colorlet{colorvariable}{mylinkcolor}

\begin{document}


\title{Equatorial stability analysis of dust particle orbits within a charged rotating disc of dust}


\author{David Rumler}
\email[]{david.rumler@uni-jena.de}
\affiliation{Theoretisch-Physikalisches Institut, Friedrich-Schiller-Universität Jena, Max-Wien-Platz 1, D-07743 Jena, Germany}


\date{\today}

\begin{abstract}
	Stability of circular orbits of the dust particles within a charged rotating disc of dust with respect to perturbations in the equatorial plane is analyzed.  Within Einstein-Maxwell theory, the charged rotating disc of dust is an axisymmetric, stationary solution given in terms of a post-Newtonian expansion. The disc solution is characterized by a rigid rotation around the axis of symmetry and a constant specific charge $\epsilon$. It is found that for $\epsilon<1$ all realized dust particle orbits are stable and for $\epsilon=1$ all dust particles are in marginally stable states.
\end{abstract}


\maketitle

\section{Introduction}
\label{sec:introduction}

An essential property of physical solutions is their stability.
If a mathematically constructed stationary solution within general relativity that is meant to describe a physical body (such as an astrophysical object) is unstable with respect to perturbations, it is usually not physically relevant.\footnote{In certain cases, a gravitational collapse caused by a perturbation is to be expected and does not imply that the solution is unphysical.}

As a physically well-defined solution of the Einstein-Maxwell equations, the charged rotating disc of dust serves as an eminent candidate to study its stability; specifically, the equatorial stability of dust particle orbits within the disc.
The axisymmetric, stationary disc solution is given in terms of a post-Newtonian expansion up to tenth order and it is characterized by a constant specific charge as well as a rigid rotation around the axis of symmetry \cite{Palenta_2013,Breithaupt_2015}.

A general investigation of circular orbits of neutral test particles in the equatorial plane of an isolated, axisymmetric, stationary, reflection symmetric, charged, rotating body, as well as a specialization on the spacetime of the charged rotating disc of dust, is provided in \cite{RumlerCircularMotion}. Circular motion of charged test particles in this general spacetime is discussed in \cite{RumlerDiss}. 

General formulae in \cite{RumlerCircularMotion} and \cite{RumlerDiss} are derived for the exterior spacetime, only. However, in case of the charged rotating disc of dust, the metric can globally be written in terms of Weyl-Lewis-Papapetrou coordinates and therefore the general formulae also hold for the interior of the disc. The formulae for charged test particles can then be applied to the charged dust particles within the disc themselves.

This allows for a detailed discussion of stability of the charged dust particle orbits within the charged rotating disc of dust.
In the present paper the stability of these circular orbits with respect to perturbations in the equatorial plane of the disc is studied in dependence of the disc's specific charge and a relativity parameter (the expansion parameter of the post-Newtonian expansion). An equatorial stability analysis of individual dust particle orbits, of course, does not represent a full analysis of stability of the disc. Nevertheless, it serves as a necessary (as well as an important) condition for stability that, compared to a full analysis, can actually be studied analytically.

In case of a vanishing specific charge an exact disc solution in terms of hyperelliptic theta functions was derived by Neugebauer and Meinel \cite{Neugebauer:1993ct, PhysRevLett.75.3046, RFE}. This uncharged rotating disc of dust can be regarded as a basic model for a disc galaxy. Meinel, Kleinwächter \cite{Meinel1995dragging} (see also \cite{RFE}) and Ansorg \cite{Ansorg1998} investigated circular geodesic motion of neutral test particles in the spacetime of the uncharged disc. As uncharged dust particles travel on geodesics, the results for the neutral test particles (moving along prograde orbits) in \cite{Ansorg1998} can be applied directly to the dust particles within the uncharged disc. With this in mind, it can be concluded that all orbits followed by the dust particles within the uncharged disc are stable.\footnote{For a stability analysis of circular orbits of neutral test particles in the equatorial plane of an axisymmetric, stationary spacetime, see, e.g., \cite{Bardeen_StableOrbits, Beheshti}.
Equatorial circular motion, including stability examinations, of neutral test particles, around Kerr and Kerr-Newman black holes, and of charged test particles, in the Reissner-Nordström and Kerr-Newman spacetime, is studied, e.g., in \cite{Bardeen1972,dadhich_equatorial_1977} and \cite{Pugliese,Balek1989II}.}

This paper is structured as follows. In \cref{sec:disc} the model of the charged rotating disc of dust is introduced and the corresponding solution in terms of a post-Newtonian expansion is presented. A detailed equatorial stability analysis of the dust particle orbits within the disc is carried out in \cref{sec:stability} by means of a stabilty condition and a related effective potential. 
Finally, \cref{sec:conclusions} provides conclusions and an outlook.

\section{Charged rotating disc of dust}
\label{sec:disc}

In the framework of Einstein-Maxwell theory an equilibrium configuration of a rigidly rotating, infinitesimally thin disc of constantly charged dust is considered. (Dust is defined as a perfect fluid with vanishing pressure.) Each dust particle within the disc carries a constant specific charge $\epsilon\in\left[0,1\right]$ (electric charge density over baryonic mass density) and travels along a circular orbit, with the constant angular velocity $\Omega$, around the axis of symmetry.\footnote{Without loss of generality the disc model is restricted to positive charges.} The corresponding spacetime is axisymmetric, stationary and reflection symmetric with respect to the equatorial plane of the disc \cite{Palenta_2013, Breithaupt_2015, Meinel_2015}. 

Not only the exterior, but the global spacetime of the charged rotating disc of dust can be characterized by the line element 
\begin{equation}\label{eq:bvp.4}
	\mathrm{d}s^2 = f^{-1}\left[ h\left( \mathrm{d}\rho^2 + \mathrm{d}\zeta^2 \right) + \rho^2\mathrm{d}\varphi^2 \right] - f\left( \mathrm{d}t + a\, \mathrm{d}\varphi \right)^2 \,,
\end{equation}
expressed in terms of Weyl-Lewis-Papapetrou coordinates (using axisymmetry and stationarity).\footnote{In this paper, $c=G=4\pi\epsilon_{0}=1$ is used.}
The four-potential of the electromagnetic field is given by
\begin{equation}
	A_a = (0,0,A_{\varphi}, A_t)\,.
\end{equation}

Utilizing the available symmetries, the disc problem, which is based on the coupled Einstein-Maxwell equations, can be reduced to a well-defined boundary value problem to the Ernst equations \cite{PhysRev.168.1415}. It was solved in terms of a post-Newtonian expansion up to tenth order by Palenta, Meinel and Breithaupt \cite{Palenta_2013,Breithaupt_2015}:
\begin{align}
	&f = 1 + \sum_{k=1}^{10}f_{2k}g^{2k} \,, \quad h = 1 + \sum_{k=2}^{10}h_{2k}g^{2k} \,, \label{eq:pne1}\\
	&a^* = \sum_{k=1}^{10}a^*_{2k+1}g^{2k+1} \,, \label{eq:pne2}\\
	&A_{\varphi}^* = \sum_{k=1}^{10}A_{\varphi\,2k+1}^{*}g^{2k+1} \,, \quad A_{t} = \sum_{k=1}^{10}A_{t\,2k}g^{2k} \label{eq:pne3}\,.
\end{align}
$a^*=\frac{a}{\rho_{0}}$ and $A_{\varphi}^{*}=\frac{A_{\varphi}}{\rho_{0}}$ are made dimensionless by normalization using the disc's coordinate radius $\rho_{0}$.
Also the angular velocity $\Omega$ (where $\Omega^*=\Omega \rho_{0}$) is given in terms of a post-Newtonian expansion:
\begin{equation}
	\Omega^{*} = \sum_{k=0}^{9}\Omega^*_{2k+1}g^{2k+1} \,.
\end{equation}
The expansion parameter $g\in\left[0,1\right]$, referred to as relativity parameter, is defined by 
\begin{equation}
	g^{2} \coloneqq 1 - \sqrt{f_{c}} \,, \quad \text{with} \quad f_{c} \coloneqq f\left(\rho=0, \zeta=0\right) \,.
\end{equation}
Alternatively, $g$ can be written by means of the redshift, $z_{c}$, gained by a photon traveling from the center of the disc ($\rho=0$, $\zeta=0$) to infinity ($\rho^2+\zeta^2 \to \infty$):
\begin{equation}
	g^2 = \frac{z_{c}}{1+z_{c}} \,.
\end{equation}
Note that $f_{2k}$, $h_{2k}$, $a^*_{2k+1}$, ... are functions of the elliptic coordinates $\eta\in[-1,1]$ and \mbox{$\nu\in[0,\infty]$} only, which are related to $\rho$ and $\zeta$ via
\begin{equation}
	\rho = \rho_0\sqrt{\left(1-\eta^2\right)(1+\nu^2)} \,, \quad \zeta=\rho_0\eta\nu \,.
\end{equation}
($\Omega^*_{2k+1}$ is independent of $\eta$ and $\nu$, as $\Omega$ is constant.)

Since the parameter space of the disc solution is fully specified by $g\in[0,1]$, $\epsilon \in [0,1]$ and the scaling parameter $\rho_{0}$, every physical quantity describing the disc, with an appropriate normalization, depends on $g$ and $\epsilon$ only (see, e.g., \cite{Rumler_MultipoleMoments}). $g\ll1$ corresponds to a Newtonian solution and $g\to1$ represents the ultra-relativistic limit (in which black hole formation is to be expected). The specific charge parameter $\epsilon$ also directly influences the disc's rotation speed, with a static configuration for $\epsilon=1$ up to a maximally rotating one for $\epsilon=0$. This can be understood by taking the Newtonian limit, in which each dust particle is in an equilibrium state of the gravitational, electric and centrifugal force.

\section{Analysis of stability within the equatorial plane}\label{sec:stability}

Due to the rigid rotation of the charged rotating disc of dust, all charged dust particles within the disc (\mbox{$0\leq\rho\leq\rho_{0}$}, $\zeta=0$) travel along equatorial circular orbits with the constant angular velocity $\Omega$. This is ensured by the boundary condition
\begin{equation}\label{eq:cm.eomctp}
	\left[ \left(-g'_{tt}\right)^{\!1/2} - \epsilon A'_{t}\right]_{,\rho} = 0 \,,
\end{equation}
where $g_{ab}$ denotes the metric and the primes the co-rotating frame of reference defined by $\varphi'=\varphi-\Omega t$.
The equation of motion that characterizes paths of charged test particles is given by
\begin{equation}\label{eq:bvp.eomLorentz}
	\frac{\mathrm{D}u^{a}}{\mathrm{d}\tau} = \epsilon F^{ab}u_{b} \,,
\end{equation}
where $u^{a}=\frac{\mathrm{d}x^{a}}{\mathrm{d}\tau}$ is the four-velocity, $\frac{\mathrm{D}u^{a}}{\mathrm{d}\tau} = {u^{a}}_{;c}u^{c}$ and $F_{ab}=A_{b,a}-A_{a,b}$.
In case of a circular motion in the equatorial plane of the disc, the $\rho$-component of \cref{eq:bvp.eomLorentz} (being the only non-vanishing component) reduces to the boundary condition (\ref{eq:cm.eomctp}).

Since the line element can be written globally in terms of Weyl-Lewis-Papapetrou coordinates, the general formulae for equatorial circular motion of charged test particles in \cite{RumlerDiss}, valid in the exterior spacetime (i.e.\ for $\rho>\rho_{0}\geq0$, where $\rho_{0}$ is the coordinate radius of a rotating body) can also be applied to the circular orbits of the dust particles within the disc (for $0\leq\rho\leq\rho_{0}$).
From the known angular velocity $\Omega$ then the specific angular momentum $\tilde{L}$ and the specific energy $\tilde{E}$ of the charged dust particles within the disc follow immediately, \linebreak using \cite{RumlerDiss}\footnote{As test particles here are the dust particles of the disc themselves, $\epsilon_{\text{tp}}=\epsilon$ and $\Omega_{\text{tp}}=\Omega$.}:
\begin{align}
	\tilde{L} &= L + \epsilon A_{\varphi} \,, \\
	\tilde{E} &= E - \epsilon A_{t} \,,
\end{align}
with
\begin{align}
	L &= \frac{g_{\varphi\varphi}\Omega+g_{\varphi t}}{\sqrt{- g_{\varphi\varphi}\Omega^{2} -2g_{\varphi t}\Omega - g_{tt}}} \,, \\
	E &= -\frac{g_{\varphi t}\Omega+g_{tt}}{\sqrt{- g_{\varphi\varphi}\Omega^{2} -2g_{\varphi t}\Omega - g_{tt}}} \,.
\end{align}
The specific angular momentum and the specific energy are conserved due to the axisymmetric and stationary spacetime.
Explicitly, $\tilde{L}$ up to first order ($k=1$) and $\tilde{E}$ up to second order ($k=2$) read:
\begin{align}
	\frac{\tilde{L}}{R_{0}}=&\left(1-\eta^{2}\right) \sqrt{1-\epsilon^{2}}\,g \notag \\
	&+\frac{1}{8} \left(1-\eta^2 \right) \sqrt{1-\epsilon^{2}}\left(\left(7\eta^{2}-\frac{5}{3}\right) \epsilon^{2} \right.\notag \\
	&\left.-\,4 \eta^{2}-\frac{2}{3} \right) g^{3} + \mathcal{O}\left(g^{5}\right) \,,  \label{eq:normL} \\
	\tilde{E}=&\, 1-\eta^{2} \left(1-\epsilon^{2} \right)  g^{2} \notag \\
	&+\frac{1}{8} \left(1-\epsilon^{2}\right) \left(1-\eta^2\right)\left(\left(7\eta^{2}-\frac{1}{3}\right) \epsilon^{2}-4\eta^{2}\right) g^{4} \notag \\
	&+ \mathcal{O}\left(g^{6}\right) \,.
\end{align}
Here, the specific angular momentum $\tilde{L}$ is normalized by the proper disc radius $R_{0} \coloneqq \int_{0}^{\rho_{0}}\!\sqrt{g_{\rho\rho}}\,\mathrm{d}\rho$ (see also \cite{Rumler}).
$\tilde{L}$ vanishes at $\rho=0$ ($\eta=\pm1$) as well as for $\epsilon=1$ (no rotation) and $\tilde{E}=1$ for $\epsilon=1$.\footnote{Due to $\tilde{E}=1$, all dust particles in a static disc with $\epsilon=1$ are marginally bound. For $\epsilon<1$ and $0\leq\rho<\rho_{0}$ all dust particles satisfy $\tilde{E}<1$ and the corresponding orbits are thus bound.}
In the simultaneous limit $\epsilon \to 0$, $\rho \to \rho_{0}$ one recovers the results of the uncharged rotating disc of dust at the rim, $L=\frac{1}{\Omega}\frac{z_{c}}{1+z_{c}}=\frac{1}{\Omega}g^2$ and $E=1$ \cite{Meinel1995dragging}. Analogous to the uncharged disc, see \cite{Ansorg1998}, also the relation $\frac{\mathrm{d}\tilde{E}}{\mathrm{d}\tilde{L}} = \Omega = \text{const.}$ holds.

Alternatively, $\tilde{L}$ and $\tilde{E}$ can be gained from the effective potential \cite{RumlerDiss}
\begin{equation}\label{eq:cm.effectivepot2}
	\mathcal{U} \coloneqq -\frac{1}{2\rho^{2}}\left(g_{tt}L^{2} +2g_{\varphi t}LE + g_{\varphi\varphi}E^2\right) + \frac{1}{2} \,,
\end{equation}
with
\begin{equation}\label{eq:cm.effectivepot}
	\frac{1}{2}g_{\rho\rho}\left(\frac{\mathrm{d}\rho}{\mathrm{d}\tau}\right)^{\!2} + \mathcal{U} = 0 \,.
\end{equation} 
Differentiation of \cref{eq:cm.effectivepot} with respect to $\tau$ leads to the equation
\begin{equation}\label{eq:cm.eom2}
	g_{\rho\rho}\frac{\mathrm{d}^2\rho}{\mathrm{d}\tau^2} + \frac{1}{2}g_{\rho\rho,\rho}\left(\frac{\mathrm{d}\rho}{\mathrm{d}\tau}\right)^{\!2} + \mathcal{U}_{,\rho} = 0 \,,
\end{equation}
which is equivalent to the $\rho$-component of the equation of motion (\ref{eq:bvp.eomLorentz}) in case of a general (not necessarily circular) test particle motion in the equatorial plane.
As a result, the equatorial motion of a test particle can equivalently be described by a motion in the effective potential.
Note that \cref{eq:cm.effectivepot2,eq:cm.effectivepot,eq:cm.eom2} hold both for charged as well as uncharged test particles (see \cite{RumlerCircularMotion}).
The conditions for circular orbits, $\frac{\mathrm{d}\rho}{\mathrm{d}\tau}=0$ and $\frac{\mathrm{d}^2\rho}{\mathrm{d}\tau^2}=0$, translate to $\mathcal{U}=0$ and $\mathcal{U}_{,\rho}=0$.

Given $\tilde{L}=\tilde{L}(\hat{\rho})$ and $\tilde{E}=\tilde{E}(\hat{\rho})$, a circular orbit at radius $\rho=\hat{\rho}$ is stable with respect to (infinitesimally small) radial and angular perturbations if and only if the extremum of $\mathcal{U}$ at \mbox{$\rho=\hat{\rho}$} is a minimum, i.e.\ $\left(\mathcal{U}_{,\rho\rho}\Big\vert_{\tilde{L}(\hat{\rho}), \tilde{E}(\hat{\rho})}\right)\!(\hat{\rho}) > 0$ \cite{RumlerCircularMotion,RumlerDiss}.

In the following, the explicit notation $\hat{\rho}$ (indicating the fixed radial location of a circular orbit), instead of $\rho$ (the radial coordinate), is only employed if both are mentioned concurrently and a notational distinction between them is necessary, such as in $\mathcal{U}_{,\rho\rho} = \left(\mathcal{U}_{,\rho\rho}\Big\vert_{\tilde{L}(\hat{\rho}), \tilde{E}(\hat{\rho})}\right)\!(\rho)$.

By introducing the dimensionless function $\mathcal{S}$ according to 
\begin{align}\label{eq:stabilityfct}
	\frac{\rho^2}{\rho_{0}^2}\mathcal{S} 
	\coloneqq&\,\, g_{tt,\rho\rho}L^{2} + 2g_{\varphi t,\rho\rho}LE + g_{\varphi\varphi,\rho\rho}E^2 - 2 \notag \\
	&+4\epsilon\left[\right.\left(g_{\varphi t,\rho}A_{t,\rho} - g_{tt,\rho}A_{\varphi,\rho}\right)L  \notag \\
	&\qquad\,\,+ \left(g_{\varphi\varphi,\rho}A_{t,\rho} - g_{\varphi t,\rho}A_{\varphi,\rho}\right)E\left.\right] \notag \\
	&+2\epsilon\left[\right.\left(g_{\varphi t}A_{t,\rho\rho} - g_{tt}A_{\varphi,\rho\rho}\right)L  \notag \\
	&\qquad\,\,+ \left(g_{\varphi\varphi}A_{t,\rho\rho} - g_{\varphi t}A_{\varphi,\rho\rho}\right)E\left.\right] \notag \\
	&+2\epsilon^2\left[g_{tt}A_{\varphi,\rho}^{\,\,\,2} - 2g_{\varphi t}A_{\varphi,\rho}A_{t,\rho} + g_{\varphi\varphi}A_{t,\rho}^{\,\,2}\right] \,,
\end{align}
the stability condition above, $\mathcal{U}_{,\rho\rho}>0$, reads for circular orbits with $0<\rho<\rho_{0}$ \cite{RumlerDiss}:
\begin{equation}
	\mathcal{S}<0 \,.
\end{equation}
The boundary points, $\rho=0$ and $\rho=\rho_{0}$, require a special treatment.\footnote{In \cref{eq:stabilityfct}, the dimensionless factor $\frac{\rho^2}{\rho_{0}^2}$ was chosen such that for $0<\rho\leq\rho_{0}$ the relation $\mathcal{S}=-2\rho_{0}^2\,\mathcal{U}_{,\rho\rho}$ holds.} 

\begin{figure}[htb]
	\centering
	\begin{minipage}{0.45\textwidth}
		\centering
		\includegraphics[scale=1]{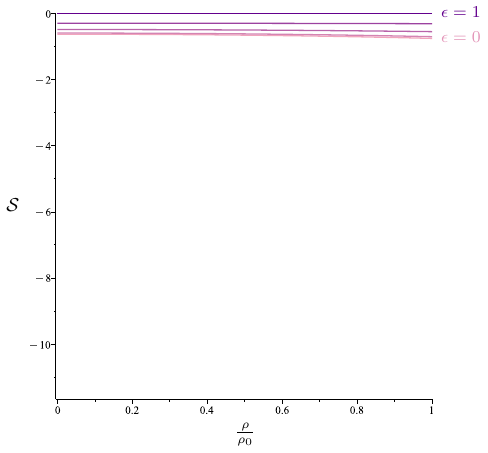}
	\end{minipage}\hfill
	\begin{minipage}{0.45\textwidth}
		\centering
		\includegraphics[scale=1]{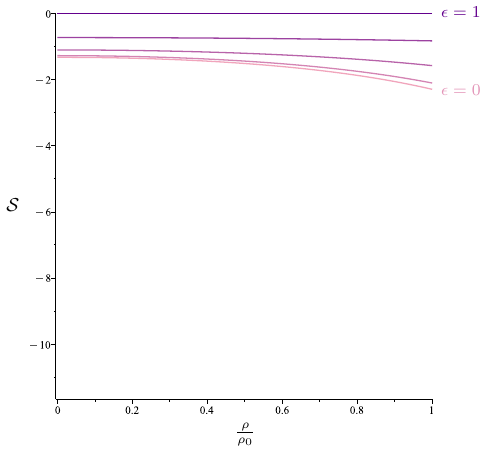}
	\end{minipage}
	\caption{$\mathcal{S}$ with $g=0.3$ (top) and $g=0.5$ (bottom), shown for \mbox{$\epsilon\in\{0, 1/4, 1/2, 3/4, 1\}$}.}
	\label{fig:1}
\end{figure}

\begin{figure}[htb]
	\centering
	\begin{minipage}{0.45\textwidth}
		\centering
		\includegraphics[scale=1]{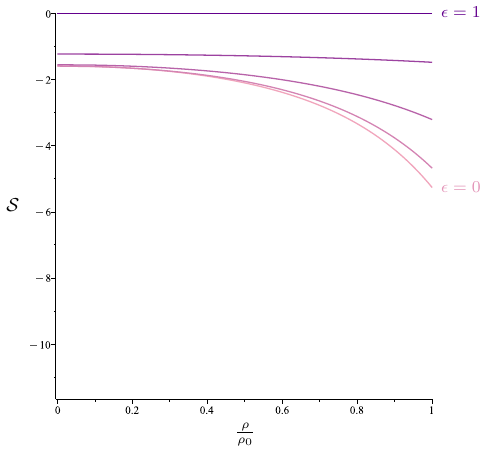}
	\end{minipage}\hfill
	\begin{minipage}{0.45\textwidth}
		\centering
		\includegraphics[scale=1]{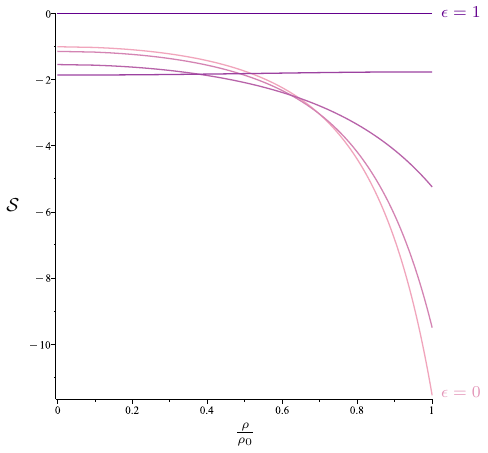}
	\end{minipage}
	\caption{$\mathcal{S}$ with $g=0.7$ (top) and $g=0.9$ (bottom), shown for \mbox{$\epsilon\in\{0, 1/4, 1/2, 3/4, 1\}$}.}
	\label{fig:2}
\end{figure}

Using the series expansions of $\tilde{L}$, $\tilde{E}$, $A_{\varphi}$ and $A_{t}$, as well as the metric $g_{ab}$, the functions $\mathcal{S}$, $\mathcal{U}$ and $\mathcal{U}_{,\rho\rho}$ can also be expressed in terms of a post-Newtonian expansion in $g$.

In \cref{fig:1,fig:2} $\mathcal{S}$ is plotted as a function of $\frac{\rho}{\rho_{0}}$ for different values of the specific charge $\epsilon$ and the relativity parameter $g$.
As can be seen, for $0<\rho<\rho_{0}$, $\epsilon=1$ leads to marginal stability (i.e.\ $\mathcal{S}=0$ or $\mathcal{U}_{,\rho\rho}=0$) and $\epsilon<1$ results in stable orbits. In the trivial case $g=0$, i.e.\ empty Minkowski spacetime, $\mathcal{S}=0$.

\begin{figure}[htb]
	\centering
	\begin{minipage}{0.45\textwidth}
		\centering
		\includegraphics[scale=1]{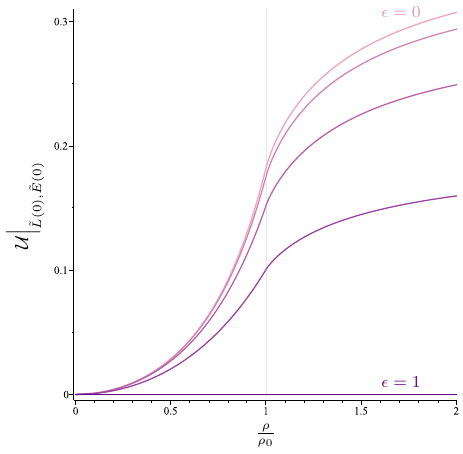}
	\end{minipage}\hfill
	\begin{minipage}{0.45\textwidth}
		\centering
		\includegraphics[scale=1]{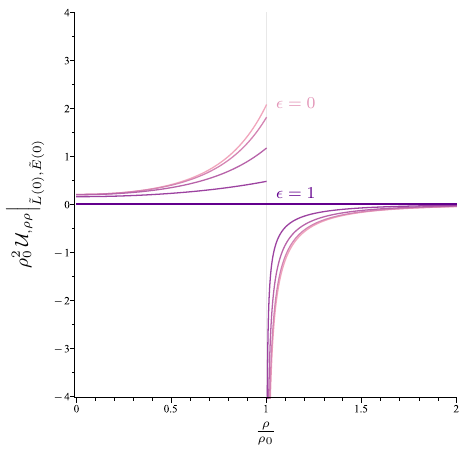}
	\end{minipage}
	\caption{$\mathcal{U}$ (top) and $\rho_{0}^2\,\mathcal{U}_{,\rho\rho}$ (bottom), with $\tilde{L}=\tilde{L}(0)$ and $\tilde{E}=\tilde{E}(0)$, shown for $g=0.7$ and \mbox{$\epsilon\in\{0, 1/4, 1/2, 3/4, 1\}$}.}
	\label{fig:3}
\end{figure}

\begin{figure}[htb]
	\centering
	\begin{minipage}{0.45\textwidth}
		\centering
		\includegraphics[scale=1]{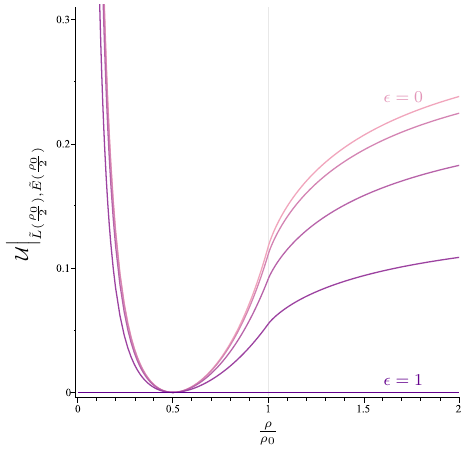}
	\end{minipage}\hfill
	\begin{minipage}{0.45\textwidth}
		\centering
		\includegraphics[scale=1]{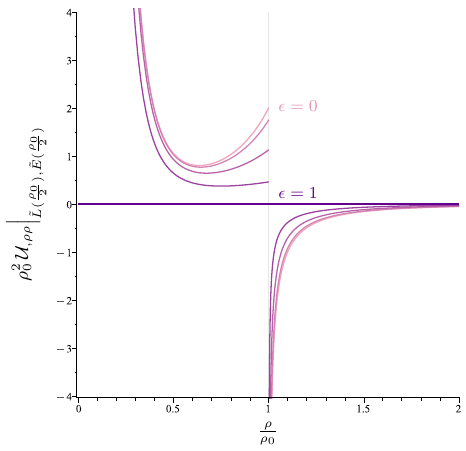}
	\end{minipage}
	\caption{$\mathcal{U}$ (top) and $\rho_{0}^2\,\mathcal{U}_{,\rho\rho}$ (bottom), with $\tilde{L}=\tilde{L}(\frac{\rho_{0}}{2})$ and $\tilde{E}=\tilde{E}(\frac{\rho_{0}}{2})$, shown for $g=0.7$ and \mbox{$\epsilon\in\{0, 1/4, 1/2, 3/4, 1\}$}.}
	\label{fig:4}
\end{figure}

\begin{figure}[htb]
	\centering
	\begin{minipage}{0.45\textwidth}
		\centering
		\includegraphics[scale=1]{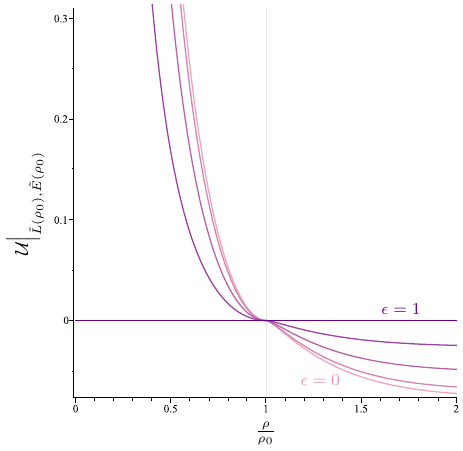}
	\end{minipage}\hfill
	\begin{minipage}{0.45\textwidth}
		\centering
		\includegraphics[scale=1]{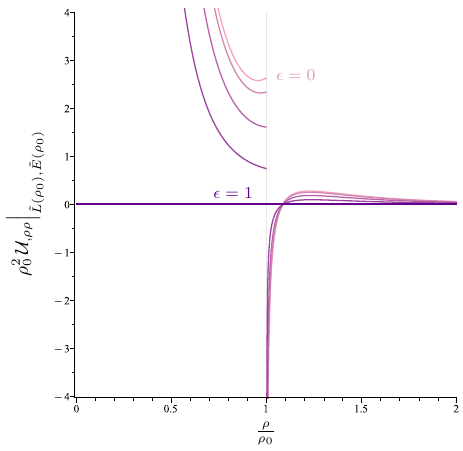}
	\end{minipage}
	\caption{$\mathcal{U}$ (top) and $\rho_{0}^2\,\mathcal{U}_{,\rho\rho}$ (bottom), with $\tilde{L}=\tilde{L}(\rho_{0})$ and $\tilde{E}=\tilde{E}(\rho_{0})$, shown for $g=0.7$ and \mbox{$\epsilon\in\{0, 1/4, 1/2, 3/4, 1\}$}.}
	\label{fig:5}
\end{figure}

With \cref{fig:3,fig:4,fig:5}, displaying the effective potential $\mathcal{U}$ and its second derivative $\mathcal{U}_{,\rho\rho}$, the boundary points $\rho=0$ and $\rho=\rho_{0}$ can be examined.

A first observation is that in the case $\epsilon=1$ for all radii $\rho\in\left[0,\rho_{0}\right]$, as well as all $\hat{\rho}\in\left[0,\rho_{0}\right]$ (with $\tilde{L}=\tilde{L}(\hat{\rho})$ and $\tilde{E}=\tilde{E}(\hat{\rho})$), $\mathcal{U}=0$ and $\mathcal{U}_{,\rho\rho}=0$. Thus, for $0\leq\hat{\rho}\leq\rho_{0}$ (non-rotating) dust particles with $\epsilon=1$ are in marginally stable states.

As shown in \cref{fig:3}, for $\epsilon<1$ the effective potential $\mathcal{U}$, with $\tilde{L}=\tilde{L}(0)$ and $\tilde{E}=\tilde{E}(0)$, has a minimum at $\rho=0$. The state of rest of the dust particle in the center of the rotating disc (with $\epsilon<1$) is therefore stable with respect to perturbations in the equatorial plane of the disc. 
Note that for $0<\rho\leq\rho_{0}$ the relation $\mathcal{S}(\rho)=-2\rho_{0}^2\,\left(\mathcal{U}_{,\rho\rho}\Big\vert_{\tilde{L}(\rho), \tilde{E}(\rho)}\right)\!(\rho)$ holds. However, at $\rho=0$ this is no longer the case (for $\epsilon<1$), since the order of limits plays a role here. $\mathcal{U}_{,\rho\rho}$, in contrast to $\mathcal{S}$, is first evaluated at $\tilde{L}=\tilde{L}(0)$ and $\tilde{E}=\tilde{E}(0)$ and afterwards the limit $\rho\to0$ is taken. 
This is a manifestation of the fact that actual circular orbits for $0<\rho\leq\rho_{0}$ (and $\epsilon<1$), with non-vanishing angular momentum, are physically distinct from a state of rest at $\rho=0$, with $L=0$. Particularly, the angular momentum prevents a dust particle at $0<\rho\leq\rho_{0}$, being exposed to a perturbation, from reaching the center. (Mathematically speaking, a circle is topologically different from a point.)

Examining the proper surface mass density of the charged rotating disc of dust, $\sigma_{\text{p}}$,\footnote{The proper surface mass density is derived from the metric functions $f=e^{2U}$ and $h=e^{2k}$: $\sigma_{\text{p}}=\frac{1}{2\pi}e^{U-k}\lim_{\zeta\to0^{+}}U'_{,\zeta}$.} reveals that (for all $\epsilon$) it vanishes at the rim, while it is non-zero everywhere else within the disc (see \cite{Palenta_2013,RumlerDiss}). Therefore, no dust particle actually travels along the orbit at the rim. 
Nevertheless, it is worth analyzing the stability of the circular orbit at the rim for $\epsilon<1$.
Taking the limit of $\mathcal{U}_{,\rho\rho}$, evaluated at $\tilde{L}=\tilde{L}(\rho_{0})$ and $\tilde{E}=\tilde{E}(\rho_{0})$, from the left- and the right-hand side of $\rho=\rho_{0}$ shows a disagreement between these limits: $\lim_{\rho\to{\rho_{0}}^{\!-}}\mathcal{U}_{,\rho\rho}\Big\vert_{\tilde{L}(\rho_{0}), \tilde{E}(\rho_{0})}>0$, whereas $\lim_{\rho\to{\rho_{0}}^{\!+}}\mathcal{U}_{,\rho\rho}\Big\vert_{\tilde{L}(\rho_{0}), \tilde{E}(\rho_{0})}=-\infty$ (see \cref{fig:5}). Therefore, $\mathcal{U}\Big\vert_{\tilde{L}(\rho_{0}), \tilde{E}(\rho_{0})}$ does not have an extremum (and thus also not a minimum) at $\rho=\rho_{0}$. The discontinuity of $\mathcal{U}_{,\rho\rho}\Big\vert_{\tilde{L}(\rho_{0}), \tilde{E}(\rho_{0})}$ -- and more generally of $\mathcal{U}_{,\rho\rho}\Big\vert_{\tilde{L}(\hat{\rho}), \tilde{E}(\hat{\rho})}$ for $0\leq\hat{\rho}\leq\rho_{0}$ (see \cref{fig:3,fig:4,fig:5}) -- at $\rho=\rho_{0}$ is not surprising, since $g_{ij,\rho\rho}$ and $A_{i,\rho\rho}$ are discontinuous  at the rim of the disc, due to the transition from the interior to the exterior disc spacetime.\footnote{Similarly, $g_{ij,\zeta}$ and $A_{i,\zeta}$ are discontinuous at the layer of the disc. The reflection symmetric disc problem yields: $\lim_{\zeta\to0^{+}}g_{ij,\zeta}=-\lim_{\zeta\to0^{-}}g_{ij,\zeta}$ and $\lim_{\zeta\to0^{+}}A_{i,\zeta}=-\lim_{\zeta\to0^{-}}A_{i,\zeta}$.}
As the orbit at the rim is neither stable (i.e.\ $\mathcal{U}_{,\rho\rho}\Big\vert_{\tilde{L}(\rho_{0}), \tilde{E}(\rho_{0})}>0$) nor unstable (i.e.\ $\mathcal{U}_{,\rho\rho}\Big\vert_{\tilde{L}(\rho_{0}), \tilde{E}(\rho_{0})}<0$), it could be classified as marginally stable. However, even if a test particle at the rim, with $\epsilon_{\text{tp}}=\epsilon$ and $\Omega_{\text{tp}}=\Omega$, were pushed towards the interior by an infinitesimally small radial perturbation, the test particle would ``roll'' back in the effective potential and disappear into the exterior spacetime.
Therefore, the circular orbit at the rim should be regarded as unstable.

In conclusion, we can state (within the given accuracy of the post-Newtonian expansion):
\begin{itemize}
	\item $\epsilon=1$ leads to marginal stability for all $\rho\in\left[0,\rho_{0}\right]$,
	\item $\epsilon<1$ results in stable orbits for $0\leq\rho<\rho_{0}$ and an unstable orbit at $\rho=\rho_{0}$.
\end{itemize}
($g=0$ is a trivial case of empty Minkowski spacetime with $\mathcal{U}=0$ and $\mathcal{U}_{,\rho\rho}=0$.)

As outlined, the orbit at the rim is not adopted by dust particles of the disc, since the proper surface mass density vanishes here. This orbit, furthermore, turns out to be unstable (for $\epsilon<1$).
In this sense, the charged rotating disc of dust (with $\epsilon<1$) thus maximizes its size by realizing all stable orbits.\footnote{The proper surface mass density of the charged rotating disc of dust with $\epsilon=1$ is determined as limit based on the full solution for all $\epsilon$. A vanishing proper surface mass density at the rim for $\epsilon=1$ is therefore to be expected. (In contrast, general configurations with $\epsilon=1$ (electrically counterpoised dust) have a proper surface mass density that is free to choose and not predetermined by their solution.)}

There is also an agreement with the uncharged rotating disc of dust: Circular geodesic orbits for $0\leq\rho<\rho_{0}$ are stable \cite{Ansorg1998} and at $\rho=\rho_{0}$ the proper surface mass density is zero \cite{Neugebauer1994,1996Neugebauer}. (In \cite{Ansorg1998}, the orbit at the rim is classified as marginally stable. However, as argued here, it is more appropriate to characterize it as unstable.)

Marginal stability of the dust particles within a static disc with $\epsilon=1$ is to be anticipated. Independent of their location, they all have $\tilde{L}=0$ and $\tilde{E}=1$. As a result, a particle at a specific location is also automatically at an equilibrium state everywhere else within the disc. So there is neither a mechanism to return the displaced dust particle to its initial position, nor can a perturbation grow.\footnote{If the perturbation is such that the dust particle, with $\epsilon=1$, is given a constant initial velocity, it would continue to move at this velocity and leave the disc.}

\begin{figure}[tb]
	\centering
		\includegraphics[scale=1]{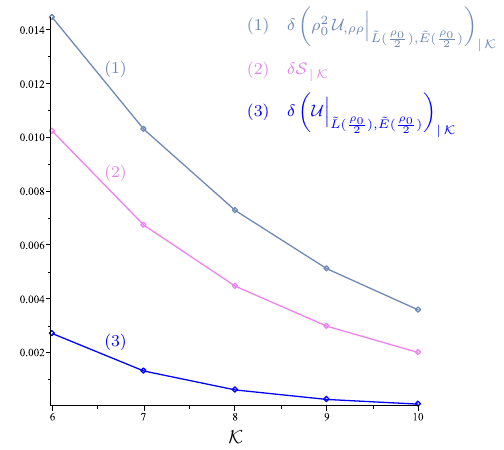}
	\caption{Estimation of the convergence behavior of $\mathcal{S}$, $\mathcal{U}\Big\vert_{\tilde{L}(\frac{\rho_{0}}{2}), \tilde{E}(\frac{\rho_{0}}{2})}$ and $\rho_{0}^2\,\mathcal{U}_{,\rho\rho}\Big\vert_{\tilde{L}(\frac{\rho_{0}}{2}), \tilde{E}(\frac{\rho_{0}}{2})}$, displayed for $g=0.9$, $\epsilon=3/4$ and $\frac{\rho}{\rho_{0}}=0.8$.}
	\label{fig:Conv}
\end{figure}

The restriction of the above discussion to $g\leq0.9$ is based on the convergence behavior of the post-Newtonian expansions of $\mathcal{S}$, $\mathcal{U}$ and $\mathcal{U}_{,\rho\rho}$. 
In order to estimate the convergence behavior of, e.g., $\mathcal{S}$ (and analogously of  $\mathcal{U}$ and $\mathcal{U}_{,\rho\rho}$), the relative change by adding a current order of the post-Newtonian expansion is computed for all orders $\mathcal{K}\leq10$:
\begin{align}\label{eq:conv}
	&\delta \mathcal{S}_{\,\vert\,\mathcal{K}} \coloneqq \frac{\left\vert \mathcal{S}_{\,\vert\,\mathcal{K}} - \mathcal{S}_{\,\vert\,\mathcal{K}-1}\right\vert}{\left\vert \mathcal{S}_{\,\vert\,\mathcal{K}}\right\vert} \,, \\
	&\text{where} \quad \mathcal{S}_{\,\vert\,\mathcal{K}} \coloneqq \sum_{k=1}^{\mathcal{K}}\mathcal{S}_{2k}g^{2k} \,.
\end{align}
As can be seen in \cref{fig:Conv}, even at $g=0.9$, $\mathcal{S}$, $\mathcal{U}$ and $\mathcal{U}_{,\rho\rho}$ exhibit good convergence behavior.

\section{Conclusions and outlook}
\label{sec:conclusions}

Dust particles within the rigidly rotating disc of charged dust all travel with the constant angular velocity $\Omega$ on circular orbits. The stability of these orbits with respect to perturbations in the equatorial plane of the disc solely depends on the specific charge $\epsilon$ and therefore on the state of rotation of the disc. In case of a static disc, for $\epsilon=1$, all dust particles are in a marginally stable state of rest. If the disc is rotating, i.e.\ $\epsilon<1$, all orbits are stable, except for the orbit at the rim of the disc, which is in fact unstable. However, none of the dust particles actually move along the orbit at the rim, since the proper surface mass density of the disc vanishes here. The charged rotating disc of dust (with $\epsilon<1$) therefore maximizes its size by occupying all stable orbits.

Another interesting question worth investigating is the stability of the dust particle orbits with respect to perturbations in $\zeta$-direction. In this case, the motion is no longer restricted to the equatorial plane of the disc. By generalizing \cref{eq:cm.effectivepot}, the motion in $\zeta$-direction can be incorporated and the corresponding stability of the dust particle orbits can be analyzed. 

Collective perturbations within the whole disc that affect more than a single dust particle orbit alter the disc's spacetime. A stability analysis of the disc against such perturbations (most certainly) requires a numerical treatment.

All orbits within the disc being equatorially stable, or at least marginally stable in case of $\epsilon=1$, underlines -- in this particular regard -- that the charged rotating disc of dust is indeed a physically meaningful solution of the Einstein-Maxwell equations.

\begin{acknowledgments}
This work has been funded by the Deutsche
Forschungsgemeinschaft (DFG) under Grant No.
406116891 within the Research Training Group RTG
2522/1. 
The author would like to thank Reinhard Meinel for valuable discussions
and Martin Breithaupt for the provided disc solution in terms of the post-Newtonian expansion up to tenth order.\\
\\
\end{acknowledgments}

\section*{Data availability}

The data that support the findings of this article are not
publicly available. The data are available from the author
upon reasonable request.



\bibliography{stability-bibliography}

\end{document}